\documentclass[useAMS]{mn2e}
\input{epsf}
\usepackage{graphicx}
\usepackage{lscape}
\usepackage{times}

\def\mnras{MNRAS}

\def\apjl{ApJL}

\def\la{\mathrel{\hbox{\rlap{\hbox{\lower4pt\hbox{$\sim$}}}{\raise2pt\hbox{$<$}}}}}
\def\ga{\mathrel{\hbox{\rlap{\hbox{\lower4pt\hbox{$\sim$}}}{\raise2pt\hbox{$>$}}}}}

\title[The ULX population of NGC 4485/4490]
{The ultraluminous X-ray source population of NGC 4485/4490.}

\author[J. C. Gladstone, T. P. Roberts]
{Jeanette C. Gladstone$^*$ and Timothy P. Roberts\\
Department of Physics, University of Durham, South Road, Durham DH1 3LE, UK\\
$^{*}$j.c.gladstone@durham.ac.uk}

\date{Submitted to MNRAS}
\pagerange{\pageref{firstpage}--\pageref{lastpage}}
\pubyear{2008}

\begin{document}

\topmargin = -0.5cm

\maketitle
\label{firstpage}

\begin{abstract} 


We report the results of spectral and temporal variability studies of the ultraluminous X-ray sources (ULXs) contained within the interacting pair of galaxies NGC 4485/4490, combining {\it Chandra} and {\it XMM-Newton} observations. Each of the four separate observations provide at least modest quality spectra and light curves for each of the six previously identified ULXs in this system; we also note the presence of a new transient ULX in the most recent observation. No short-term variability was observed for any ULX within our sample, but three out of five sources show correlated flux/spectral changes over longer timescales, with two others remaining stable in spectrum and luminosity over a period of at least five years. We model the spectra with simple power-law and multi-colour disc black body models. Although the data is insufficient to statistically distinguish models in each epoch, those better modelled (in terms of their $\chi^2$ fit) by a multi-colour disc black body appear to show a disc-like correlation between luminosity and temperature, whereas those modelled by a power-law veer sharply away from such a relationship. The ULXs with possible correlated flux/spectral changes appear to change spectral form at $\sim 2 \times 10^{39}$  erg s$^{-1}$, suggestive of a possible change in spectral state at high luminosities. If this transition is occuring between the very high state and a super-Eddington ultraluminous state, it indicates that the mass of the black holes in these ULXs is around 10--15 $M_\odot$. 

\end{abstract}


\section{Introduction} 

Ultraluminous X-ray Sources (ULXs) are point like X-ray sources
situated external to the nucleus of their host galaxy, which have
inferred X-ray luminosities in excess of 10$^{39}$ erg
s$^{-1}$. Although, first observed $\sim$30 years ago by {\it
Einstein} (Fabbiano 1989), these sources are yet to be fully
understood, with the only certainty being that they cannot all be
explained by stellar-mass black holes (with masses 3--20 $M_\odot$)
radiating isotropically below their Eddington limit. The simplest way
to circumvent this issue is to assume that larger compact objects
reside within these systems, an approach that would provide us with
objects intermediate in mass (IMBH, of mass $\sim$100--10,000
$M_\odot$; Colbert \& Mushotsky 1999), as well as luminosity, between
those of their stellar-mass and super-massive black hole cousins. This
idea is appealing as it can provide a `missing link' in the mass scale
of these objects and would have significant implications on our
understanding of the formation and evolution of both black holes and
galaxies (e.g. Madau and Rees 2001; Ebisuzaki et al. 2001). Evidence
supporting this scenario was presented in 2003 by Miller et al. in
which the spectra of NGC 1313 X-1 and X-2 were fitted with a standard
canonical accretion disc plus power-law continuum model for accreting
black hole X-ray binary systems. This resulted in a remarkably cool
disc temperature of kT$\simeq$150 eV, that implies a black hole mass
of $\sim 2\times 10^{4}$ $M_\odot$. Similar results were found for
other sources (e.g. Miller et al 2004a \& 2004b). However recent
studies, such as Stobbart, Roberts \& Wilms (2006), Gon{\c c}alves \&
Soria (2006), Vierdayanti et al. (2006) and Feng \& Kaaret (2007) have
demonstrated that the spectra of ULXs can be equally well explained by
models that infer a stellar-mass black hole (perhaps up to
100 $M_\odot$) accreting at extreme rates to be powering the system.

To date, X-ray spectral analysis has been unable to uniquely determine
the nature of ULXs. One option is to turn to a broader bandpass in an
attempt to find counterparts to these sources. Alternatively, others
have chosen to carry out X-ray variability studies on these highly
luminous systems, to compare to the known variability characteristics
of stellar-mass sources and AGN on both short and long timescales. The
analysis of temporal variability for both stellar and super-massive
black holes via power spectral densities (PSD) has shown a scaling by
mass and accretion rates of the PSD break time-scales (e.g. McHardy et
al 2006). Unfortunately very few ULXs show much variability power, but
some measurements have been made, for example Soria et al (2004)
detected a break at 2.5 mHz in the PSD of NGC 5408 X-1, which
indicated a mass of $\sim$100 $M_\odot$. In fact, intra-observational
studies have revealed that only $\la$15\% of ULXs exhibit measurable
short-term variability (Swartz et al 2004; Feng \& Kaaret 2005).

When extending variability studies to longer timescales problems arise
because few sources have been regularly observed. Unlike their
Galactic stellar-mass counterparts or the brighter quasars, which can
be monitored daily by instruments such as the All Sky Monitor on board
the {\it Rossi X-ray Timing Explorer}, analysis can only be carried
out using pointed observations. This leads to large gaps in the light
curves, making it difficult to track the trends of these
systems. However, findings to date show that ULXs tend to be
persistent over timescales of $\ga$10 years (e.g. Roberts et
al. 2004). One avenue that remains under-explored is that of spectral
variability. These studies suffer from the same windowing issues
discussed above, making it difficult to observe spectral evolution in
detail. Some studies have been performed, such as the investigation
into the X-ray source population of the Antennae performed by Fabbiano
et al (2003). This, along with other studies, found that the majority
of ULXs display a general hardening as the luminosity of the system
increases (e.g. Homberg II X-1, Dewangan et al. in 2004; NGC 5204 X-1,
Roberts et al. 2006).

In 2002, Roberts et al. published results from a 20 ks {\it Chandra}
ACIS-S observation of the interacting galaxy pair NGC~4485 and NGC
4490 (this paper is referred to hereafter as RWWM02). These galaxies
reside at a distance $\sim$7.8 Mpc, placing these amongst the very
nearest interacting pairs of late type galaxies. Initially their work
entailed a study of the general X-ray emission from the pair,
considering the emission from both diffuse and discrete sources. A
total of 29 discrete sources were found to be coincident with NGC 4490
and one with NGC 4485, with luminosities ranging from
$\sim$2x10$^{37}$ to 4x10$^{39}$erg s$^{-1}$. Of these, six were
identified as emitting at ultraluminous rates, a number greatly in
excess of the average of $<$ 1 per galaxy (Liu, Bregman \& Irwin
2006), and a population of ULXs only bettered numerically by M51 and
M82 within 10 Mpc.  Here we revisit the data presented within RWWM02,
combining this with information from subsequent {\it Chandra} and {\it
XMM-Newton} observations in order to better explore the behaviour of
these ULXs on both short and long timescales.

\section{Observations and Data Reduction}

The NGC 4485/4490 galaxy pair has been observed by {\it Chandra} on
three occasions, the first of which was reported in RWWM02, with two
subsequent deeper observations taken during 2004. A search of the {\it
XMM-Newton} archive revealed one further observation taken in
2002. Details of these observations can be seen in Table
\ref{tab:telescope}. (For simplicity, the observations will be
labelled as shown in this Table throughout the rest of this paper.)

\renewcommand{\baselinestretch}{1.0} %
\begin{table*}
\leavevmode
\begin{center}
\caption{Recent X-ray observations of NGC 4485 and NGC 4490.}
\begin{tabular}{ccccccccll}
\hline
Observatory & Detector & Obs ID     & Label & Mode       & Date       & Time$^a$ & Exposure$^b$ & \multicolumn{2}{c}{Aimpoint$^{c}$} \\
            &          &            &       &            &            &          & (ks)   & \multicolumn{1}{c}{RA} & \multicolumn{1}{c}{Dec}\\
\hline
{\it Chandra \/}     & ACIS-S   & 1579       & C1    & FAINT      & 2000:11:03 & 01:24:04 & 19.7       & 12$^h$ 30$^m$ 31$^s$.2  & +41$^o$ 39' 00''  \\
{\it XMM-Newton \/}  & EPIC     & 0112280201 & X1    & Full Field & 2002:05:27 & 07:15:20 & 12.8       & 12$^h$ 30$^m$ 30$^s$.28 & +41$^o$38'53''.8 \\
{\it Chandra \/}     & ACIS-S   & 4725       & C2    & VFAINT     & 2004:07:29 & 20:44:43 & 39.0       & 12$^h$ 30$^m$ 31$^s$.2  & +41$^o$ 39' 00''  \\
{\it Chandra \/}     & ACIS-S   & 4726       & C3    & VFAINT     & 2004:11:20 & 06:30:59 & 40.1       & 12$^h$ 30$^m$ 31$^s$.2  & +41$^o$ 39' 00''  \\
\hline
\end{tabular}
\end{center}
\begin{minipage}{\textwidth}
Notes: $^a$Exposure start time in universal time (UT).  $^b$ Instrument ontime (for EPIC we quote the pn value).  $^c$J2000 coordinates. 
\end{minipage}
\label{tab:telescope}
\end{table*}

\subsection{Chandra observations} %

{\it Chandra} observations were performed with the Advanced CCD
Imaging Spectrometer (ACIS), using the S-array. The first observation
(C1) was positioned on the standard aimpoint for the back-illuminated
S3 chip, although a Y offset of -2.0 arcmins was used during
subsequent observations to centralise the galaxy pair on the S3
chip. The later observations were also performed with the telemetry in
VFAINT mode to optimise the detection of the faint extended diffuse
emission component in the galaxies detected by RWWM02.  Data reduction
was carried out using the \textsc{ciao} software, version 3.3.0.1, and
standard {\it Chandra} data analysis threads\footnote{For {\it
Chandra} threads, published by the {\it Chandra} X-ray Center, see
{\tt http://asc.harvard.edu/ciao}.}. The three observations were
filtered by energy, rejecting events with energies outside the range
0.3--10 keV. The source spectra and light curves were extracted for
each source using the \textsc{ciao} tasks {\it PSEXTRACT} and {\it
DMEXTRACT}, taking data from circular apertures, 5 pixels in diameter
(thus encircling $> 90\%$ of the source energy) and centred on each
ULX.

Background spectra were extracted via an annulus encircling the source
(8 to 16 pixels in diameter). Background light curves were extracted
from circular regions, the same size as the source region, positioned 
in a source-free region near each ULX. The response and ancillary
response files, necessary for spectral analysis, were created
automatically by the standard {\it Chandra} tasks.

\subsection{XMM-Newton observation} %

Data from the {\it XMM-Newton} observation of the NGC 4485/4490 pair was
collected from the {\it XMM-Newton} science archive (XSA) and reduced
using the \textsc{sas} software (version 7.0.0). Here we utilise data
from the {\it XMM-Newton} European Photon Imaging Camera (EPIC), which
was operated in full-field mode with a medium optical filter in place
for all 3 cameras during the observation. This observation was
affected by a background flare which was removed using a good time
intervals (GTI) file, created using a full-field background light
curve extracted from the PN camera data.  As a result of the applied
time filter the total ontime was reduced to 12.8 ks.  The same GTI
file was used for both the PN and MOS instruments to allow for the
direct co-addition of light curves. As a final check for unscreened
flares we visually inspected the time filtered full-field light curves
for each MOS detector, and no additional flares were observed.

Source spectra and light curves were extracted from circular regions
centred on the individual ULXs in each detector. We used a region 16
arcseconds in radius where possible, to enclose as many source counts
as possible, whilst minimising the contribution from other components
within the galaxies. The encircled energy fraction for such regions is
$\sim 70\%$ for both MOS and PN detectors.  Where cross-contamination
was likely due to the close proximity of these and other bright
sources, data was extracted from a smaller circular region of 8 or 12
arcseconds in radius (dropping the encircled energy fractions to $\sim
50 - 60\%$; specific details of the regions used for individual
sources are discussed below). We selected the best quality data
(\textsc{FLAG} = 0) in each case and screened with \textsc{PATTERN}
$\leq$ 4 for PN and \textsc{PATTERN} $\leq$ 12 for MOS, with light
curves extracted in the 0.3--10.0 keV band. The response and ancillary
response files were created automatically by the standard {\it
XMM-Newton} tasks.

Background spectra and light curves were extracted from circular
regions for every source. Regions were identical in size to source
data extraction regions for light curves, but larger regions were used
for spectra. Every effort was made to position these on the same chip
as the source at a similar off-axis angle and distance from the read
out nodes, though this was not always possible for the spectral
regions. When necessary additional background regions were used to
provide background data with average detector characteristics similar
to the source data.


\begin{figure*}
\leavevmode
\begin{minipage}{5cm}
\epsfxsize=5.35cm {\rotatebox{0}{\epsfbox{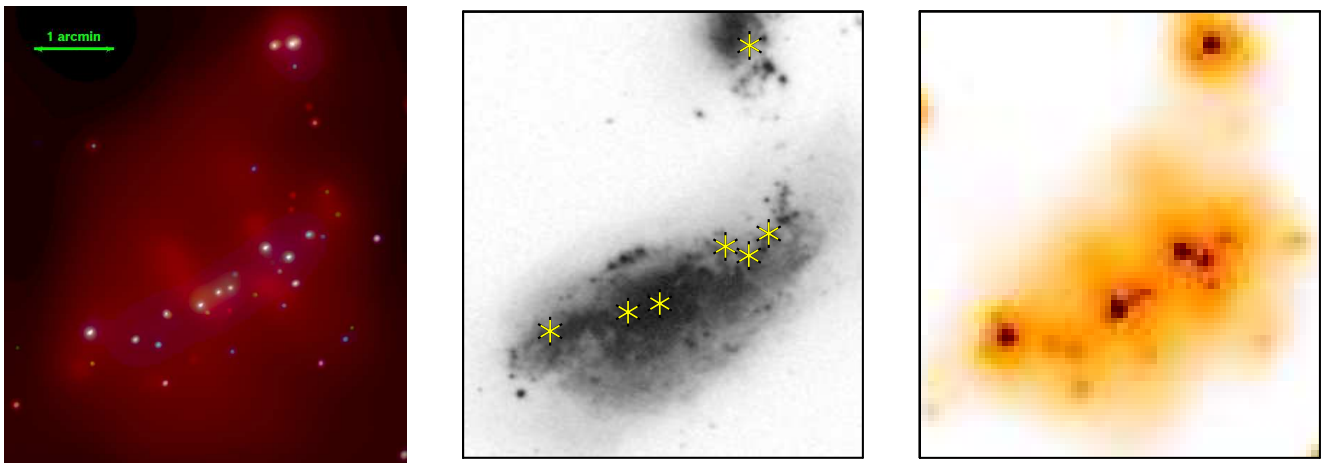}}} 
\end{minipage}\hspace*{1cm}
\begin{minipage}{5cm}
\epsfxsize=6cm {\rotatebox{-90}{\epsfbox{optical.ps}}} 
\end{minipage}\hspace*{1cm}
\begin{minipage}{5cm}
\epsfxsize=6cm{\rotatebox{-90}{\epsfbox{xmm_smoothed.ps}}}
\end{minipage}
\caption{X-ray and optical views of the galaxy pair NGC
4485/4490. Each panel is shown on the same spatial scale to facilitate
direct comparison, and North is up in all three panels.  ({\it
left\/}) True-colour {\it Chandra\/} image.  We combined data from all
three {\it Chandra\/} ACIS-S observations using standard \textsc{ciao}
threads. We combine the data using \textsc{merge} and create the image
using the DS9 thread. The data is adaptively smoothed using the
\textsc{csmooth} algorithm. The colours represent emission in the 0.3
- 1 keV (red), 1 - 2 keV (green) and 2 - 8 keV (blue) bands.  The
figure shows that the numerous luminous point sources within NGC 4490
(and NGC 4485) are clearly embedded in an extended, soft X-ray
emission component that is present in both the disc and halo regions
of the main galaxy.  ({\it centre\/}) Optical image of the galaxies.
We display the Digitised Sky Survey blue data, and mark the positions
of the seven ULXs detailed in this paper using yellow asterisks. ({\it
right\/}) {\it XMM-Newton\/} data for the galaxies in the 0.3--10.0
keV band.  The raw (i.e. not exposure-corrected) imaging data from all
three EPIC cameras is co-added, and then smoothed using
\textsc{csmooth}, before being displayed on a log heat scale.}
\label{fig:image}
\end{figure*}

\section{ULX properties} %

Our source selection was based on the `bright' sample criteria set out
by RWWM02, that is to say we selected sources that have observed X-ray
luminosities of 10$^{39}$ erg s$^{-1}$ and above (i.e. objects
ultraluminous in nature). The authors identified a total of six ULXs
within these two galaxies that are listed in Table \ref{tab:param}.
It should be noted that the nomenclature for ULXs used here follows
that of RWWM02, who used the official IAU designations for {\it
Chandra} sources incorporating the source position in J2000
coordinates. Each observation was also checked to identify any
previously unobserved ULXs that had emerged since the original study
had been carried out. One additional source was found (in observation
C3), which we refer to hereafter as CXOU J123038.3+413830 using the
above nomenclature\footnote{Fridriksson et al. 
(2008) also noted the emergence of this source and one other possible ULX, 
CXOU J123035.1+413846. We extracted a spectrum for the latter finding 
an observed luminosity of $<$10$^{39}$ erg s$^{-1}$ in the 0.3 - 10.0 
keV band, so it is not considered further in our analysis.}.  The
seven ULXs (where detected) had ACIS-S/combined EPIC count rates of
$0.009 - 0.076$/$0.018 - 0.105$ count s$^{-1}$, resulting in $290 -
2070$ counts in total being accumulated per source per observation.
We highlight their positions relative to the optical extent of the
galaxies using Digitised Sky Survey data, which we display alongside
{\it Chandra\/} and {\it XMM-Newton\/} images constructed from the
observations studied herein, in Figure~\ref{fig:image}.

The spectra and light curves for the seven sources were extracted and
analysed for each observation, providing their data was sufficiently
good to allow spectral analysis.  This was only possible for CXOU
J123038.3+413830 during observation C3. Conversely CXOU
J123029.5+413927 had sufficient data for spectral analysis in all but
the {\it XMM-Newton} observation (X1). The source spectra and light
curves for CXOU J123030.8+413911, CXOU J123029.5+413927 (light curves
only) and CXOU J123032.3 +413918 were extracted from X1 using a region
of only 8 arcseconds radius due to the close proximity of these three
sources to one another.  During the X1 observation, CXOU
J123030.6+414142 was unfortunately positioned close to a PN chip gap.
We therefore used a polygonal region to optimise our extraction of the
PN source data.  Circular apertures (radius 16 arcseconds) were viable
for this source in each MOS detector.  Similarly, PN data from CXOU
J123042.3+413818 was extracted from a 12 arcsecond radius region due
to the proximity of a chip gap, whilst its MOS data was extracted from
a full 16 arcsecond radius aperture.

\subsection{X-ray spectra} %

All the ULX X-ray spectra were grouped to a minimum of 20 counts per
bin and then fitted using simple (absorbed single component) models in
\textsc{xspec} version 11.3.1. The spectral fitting was carried out in
the 0.3--10.0 keV range for all instruments.  When fitting models to the
spectra from {\it XMM-Newton}, the PN and MOS data were analysed
concurrently to gain the best fit.  A constant multiplicative factor
was applied to allow for calibration differences between cameras. The
PN constant was fixed at unity, whilst the constants for each of the
MOS detectors were left as free parameters and, for the majority of
sources, they agree to within 10\% (larger discrepancies are only
present where disparate extraction regions were necessary; see above).

Each model included two absorption components; a fixed column of $1.81
\times 10^{20} \rm ~cm^{-2}$ representing the known absorption column
along the line of sight to NGC 4485/4490 in our own Galaxy (from the
Leiden, Argentine and Bonn Survey; Kalberia et al 2005), and a second
component left free to vary that represents absorption within the host
galaxy.  These absorption columns were modelled using the
\textsc{tbabs} model (Wilms, Allen \& McCray 2000). In our initial
fits it acted on either a power-law continuum (\textsc{po} in
\textsc{xspec} syntax) or a multi-colour disc blackbody model
(\textsc{diskbb}; Mitsuda et al. 1984), similar to RWWM02. We did
attempt more complicated models, for example two component models such
as an absorbed disc + power-law model, but no statistically
significant improvement was achieved in the fitting of any of the
data, and so they are not discussed any further in this paper.

The results of our fitting for the two models (absorbed power-law
continuum and absorbed multi-colour disc blackbody; hereafter referred
to as the PL and MCDBB models respectively) are shown in
Table~\ref{tab:param}.  This also includes a luminosity for each ULX,
the luminosity and its errors were calculated using the best fitting
model to the data. For {\it Chandra} data this calculation proved
straightforward, but for the concurrently fit {\it XMM-Newton} data
sets we estimated the model flux from the PN data only. As CXOU
J123030.6+414142 was unfortunately positioned close to a chip gap in
the PN detector, we derived its luminosity using an average of the two
MOS detectors. The errors quoted throughout Table
\ref{tab:param} are the 90 per cent errors for one interesting
parameter.

\renewcommand{\baselinestretch}{1.0} %
\begin{table*}
\leavevmode
\caption{Spectral modelling of the ULXs in NGC 4485 and NGC 4490}
\begin{center}
\begin{tabular}{ccccccccccc}
\hline
Source Name & Obs &                        & TBABS * PO    &                  &                      &                        & TBABS * DISKBB          &         &                \\
CXOU        &     & $N_H$$^a$              & $\Gamma$      & $\chi^{2}$ / DoF & Luminosity$^b$       & $N_H$$^a$              & $kT_{in}$$^c$  & $\chi^{2}$ / DoF & Luminosity$^b$ \\
\hline
            &  C1 & 5$^{+2}_{-1}$          & 3.9$^{+0.9}_{-0.8}$    &  \textbf{6.3/10}          & 0.9$^{+0.1}_{-0.8}$  & 3.0$^{+0.7}_{-0.6}$    & 0.8$^{+0.2}_{-0.1}$    & 6.4/10  & 0.86$^{+0.03}_{-0.6}$ \\
J123029.5   &  X1 & ...                    & ...                    & ...              & ...                  & ...                    & ...                    & ...              & ...           \\
   + 413927 &  C2 & 3.3$^{+1.0}_{-0.7}$    & 2.5$^{+0.6}_{-0.4}$    & \textbf{33.1/18} & 1.0$^{+0.02}_{-0.7}$ & 2.2$^*$                & 1.3$^*$                & 38.6/18	   & 0.9$^*$         \\
            &  C3 & 2.1$^*$                & 2.0$^*$                & \textbf{26.7/13} & 0.7$^*$              & 1.4$^*$                & 1.4$^*$                & 29.0/13          & 0.6$^*$                \\
            &     &                        &                        &                  &                      &                        &                        &                  &             \\
            &  C1 & 0.37$^{+0.08}_{-0.07}$ & 1.72$^{+0.07}_{-0.1}$  & 61.1/59 & 4.0$^{+0.4}_{-0.7}$  & 0.16$^{+}_{-}$0.05     & 1.4$^{+0.2}_{-0.1}$    & \textbf{46.8/59}          & 3.6$^{+0.2}_{-0.7}$  \\
J123030.6   &  X1 & 0.41$^{+}_{-}$0.1      & 2.1$^{+}_{-}$0.2      & \textbf{35.9/38}          & 2.1$^{+0.6}_{-1}$    & 0.09$^{+}_{-}$0.08     & 1.2$^{+0.2}_{-0.1}$    & 39.5/38 & 1.9$^{+0.3}_{-1}$  \\
   + 414142 &  C2 & 0.36$^{+0.7}_{-0.6}$   & 2.0$^{+}_{-}$0.1       & 84.1/66          & 2.4$^{+0.3}_{-0.4}$  & 0.14$^{+0.05}_{-0.04}$ & 1.2$^{+}_{-}$0.1       & \textbf{77.1/66} & 2.12$^{+0.07}_{-0.3}$ \\
            &  C3 & 0.40$^{+}_{-}$0.07     & 1.87$^{+0.1}_{-0.06}$  & 70.2/69 & 2.6$^{+0.3}_{-0.5}$  & 0.14$^{+}_{-}$0.05     & 1.4$^{+0.1}_{-0.10}$   & \textbf{55.3/69}          & 2.2$^{+0.1}_{-0.3}$ \\
            &     &                        &                        &                  &                      &                        &                        &                  &             \\
            &  C1 & 1.1$^{+0.4}_{-0.3}$    & 1.8$^{+}_{-}$0.3       & \textbf{24.2/29}          & 2.5$^{+0.4}_{-1}$    & 0.5$^{+}_{-}$0.2       & 1.7$^{+0.4}_{-0.3}$    & 28.8/29 & 2.3$^{+0.2}_{-2}$  \\
J123030.8   &  X1 & 1.2$^{+0.3}_{-0.2}$    & 1.8$^{+0.3}_{-0.2}$    & \textbf{28.3/32}          & 2.8$^{+0.9}_{-2}$    & 0.7$^{+}_{-}$0.2       & 1.6$^{+0.3}_{-0.2}$    & 32.2/32 & 2.5$^{+0.6}_{-3}$  \\
   + 413911 &  C2 & 2.0$^{+}_{-}$0.3       & 2.4$^{+}_{-}$0.2       & \textbf{55.9/53} & 2.6$^{+0.4}_{-1}$    & 1.2$^{+}_{-}$0.2       & 1.2$^{+}_{-}$0.1       & 60.5/53          & 2.4$^{+0.1}_{-0.5}$ \\
            &  C3 & 1.5$^{+}_{-}$0.2       & 2.2$^{+}_{-}$0.2       & \textbf{64.4/59} & 2.4$^{+0.3}_{-0.7}$  & 0.9$^{+}_{-}$0.1       & 1.3$^{+}_{-}$0.1       & 65.5/59          & 2.3$^{+0.07}_{-0.5}$ \\
            &     &                        &                        &                  &                      &                        &                        &                  &             \\
            &  C1 & 0.6$^{+}_{-}$0.2       & 1.7$^{+}_{-}$0.3       & \textbf{12.8/21}          & 1.7$^{+0.2}_{-0.6}$  & 0.3$^{+0.1}_{-0.3}$    & 1.5$^{+0.5}_{-0.2}$    & 14.2/21 & 1.5$^{+0.2}_{-1}$  \\
J123032.3   &  X1 & 1.0$^{+}_{-}$0.1       & 2.2$^{+}_{-}$0.2       & 57.4/47          & 2.8$^{+0.9}_{-1}$    & 0.5$^{+}_{-}$0.1       & 1.2$^{+}_{-}$0.1       & \textbf{47.3/47} & 2.7$^{+0.3}_{-0.9}$  \\
   + 413918 &  C2 & 1.0$^{+}_{-}$0.2       & 2.0$^{+}_{-}$0.2       & 72.6/59          & 3.2$^{+0.4}_{-1}$    & 0.5$^{+}_{-}$0.1       & 1.4$^{+0.2}_{-0.1}$    & \textbf{67.0/59} & 2.9$^{+0.1}_{-0.5}$ \\
            &  C3 & 0.9$^{+}_{-}$0.1       & 1.9$^{+}_{-}$0.1       & 88.1/64          & 2.5$^{+0.3}_{-0.5}$  & 0.53$^{+0.10}_{-0.08}$ & 1.5$^{+0.2}_{-0.1}$    & \textbf{83.8/64} & 2.4$^{+0.1}_{-0.7}$ \\
            &     &                        &                        &                  &                      &                        &                        &                  &             \\
            &  C1 & 0.6$^{+0.3}_{-0.2}$    & 1.7$^{+}_{-}$0.3       & \textbf{20.5/22}          & 1.7$^{+0.3}_{-0.1}$  & 0.3$^{+}_{-}$0.2       & 1.4$^{+0.5}_{-0.2}$    & 23.3/22 & 1.5$^{+0.1}_{-1}$  \\
J123036.3   &  X1 & 0.58$^{+0.1}_{-0.09}$  & 2.06$^{+0.1}_{-0.07}$  & 84.7/66          & 2.7$^{+0.7}_{-1}$    & 0.23$^{+0.07}_{-0.06}$ & 1.3$^{+}_{-}$0.1       & \textbf{77.3/66} & 2.6$^{+0.4}_{-0.8}$  \\
   + 413837 &  C2 & 0.8$^{+0.4}_{-0.2}$    & 2.5$^{+0.5}_{-0.3}$    & \textbf{10.7/13} & 0.40$^{+0.08}_{-0.3}$& 0.2$^{+}_{-}$0.2       & 1.0$^{+0.3}_{-0.2}$    & 18.9/13          & 0.39$^{+0.06}_{-0.2}$ \\
            &  C2 & 1.3$^{+}_{-}$0.2       & 2.5$^{+}_{-}$0.2       & 72.9/62          & 2.1$^{+0.3}_{-0.5}$  & 0.7$^{+}_{-}$0.1       & 1.03$^{+}_{-}$0.08     & \textbf{54.9/62} & 1.94$^{+0.06}_{-0.2}$ \\
            &     &                        &                        &                  &                      &                        &                        &                  &             \\
            &  C1 & ...                    & ...                    & ...              & ...                  & ...                    & ...                    & ...              & ...           \\
J123038.3   &  X1 & ...                    & ...                    & ...              & ...                  & ...                    & ...                    & ...              & ...           \\
   + 413830 &  C2 & ...                    & ...                    & ...              & ...                  & ...                    & ...                    & ...              & ...           \\
            &  C3 & 1.8$^{+0.4}_{-0.3}$    & 2.6$^{+0.3}_{-0.2}$    & 41.5/35          & 1.2$^{+0.2}_{-0.5}$  & 1.1$^{+}_{-}$0.2       & 1.1$^{+}_{-}$0.1       & \textbf{35.8/35} & 1.2$^{+0.04}_{-0.3}$  \\
            &     &                        &                        &                  &                      &                        &                        &                  &             \\
            &  C1 & 1.3$^{+}_{-}$0.2       & 2.3$^{+}_{-}$0.2       & 46.6/40          & 2.91$^{+0.4}_{-0.09}$ & 0.8$^{+0.2}_{-0.1}$    & 1.2$^{+}_{-}$0.1       & \textbf{38.6/40} & 2.7$^{+0.1}_{-0.5}$  \\
J123043.2   &  X1 & 1.1$^{+0.2}_{-0.1}$    & 2.7$^{+}_{-}$0.2       & 59.3/54          & 2.2$^{+0.6}_{-1}$    & 0.47$^{+0.1}_{-0.09}$  & 0.97$^{+0.09}_{-0.08}$ & \textbf{51.7/54} & 2.1$^{+0.3}_{-0.7}$  \\
   + 413818 &  C2 & 1.3$^{+}_{-}$0.2       & 2.7$^{+}_{-}$0.2       & 83.6/58          & 1.9$^{+0.2}_{-0.6}$  & 0.7$^{+}_{-}$0.1       & 0.94$^{+0.08}_{-0.07}$ & \textbf{68.4/58} & 1.75$^{+0.05}_{-0.2}$ \\
            &  C3 & 1.1$^{+0.2}_{-0.1}$    & 2.2$^{+}_{-}$0.1       & 87.2/79          & 3.1$^{+0.3}_{-0.6}$  & 0.58$^{+0.10}_{-0.09}$ & 1.32$^{+}_{-}$0.1      & \textbf{82.0/79} & 2.9$^{+0.1}_{-0.3}$  \\
\hline
\end{tabular}
\end{center}
\begin{minipage}{\textwidth}
Notes: the data quality was insufficient for spectral fitting in the
{\it XMM-Newton} observation of CXOU J123029.5+413927.  CXOU J123038.3
+ 413830 is the new transient source, not previously observed as a
ULX, that appears only in the most recent observation. We embolden the
$\chi^{2}$/degrees of freedom (DoF) for the best fitting model for
each source observation.  Specific notes: $^a$ Absorption column in
units of 10$^{22}$ atoms cm$^{-2}$.  $^b$ Observed luminosity in the
0.5--8.0 keV band in units of 10$^{39}$erg s$^{-1}$.  $^c$ Disc
temperature in keV. $^*$ The best fitting models to this data gave a
reduced $\chi^{2}$ greater than 2, therefore we did not place
constraints on the parameter errors.
\end{minipage}
\label{tab:param}
\end{table*}

\begin{figure}
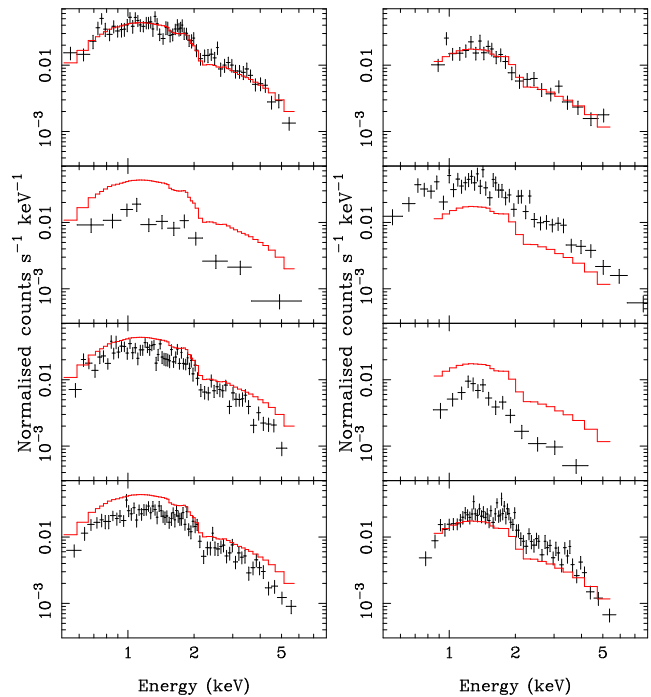

\leavevmode
\begin{minipage}{85mm}
\begin{center}
\includegraphics[height=42mm, angle=-90]{src6_tbabs_tbabs_and_po.ps} 
\includegraphics[height=42mm, angle=-90]{src7_tbabs_tbabs_and_po.ps} 
\end{center}
\caption{Example X-ray spectra from two ULXs.  We show the spectral
data for CXOU J123030.6+414142 (left) and CXOU J123036.3+413837
(right) over all four observations, which are ordered top - bottom in
chronological order (i.e. C1, X1, C2, C3).  In each panel we show the
spectral data points in black (NB. for clarity we omit EPIC-MOS data,
showing only EPIC-pn data in the second panel down).  For ease of
direct comparison we display the best fitting power-law continuum
model to the C1 data in red in all subsequent panels.  Clearly both
sources display considerable spectral and luminosity variability.}
\label{fig:spectra}
\end{minipage}
\end{figure}

The simple absorbed single component models as shown in
Table~\ref{tab:param} provide acceptable fits to the majority of the
data (model rejection probabilities, based on the $\chi^2$ fit, of
$P_{\rm rej} < 95\%$).  Only three best-fits were statistically poor
models of the data; both the latter {\it Chandra\/} observations of
CXOU J123029.5+413927 (with rejection probabilities of $P_{\rm rej} >
98.4\%$), and a very marginal rejection of the best fit to CXOU
J123032.3+413918 in observation C3 ($P_{\rm rej} = 95.1\%$).  We show
examples of the spectral quality of our data in
Figure~\ref{fig:spectra}, where data for two of the ULXs is shown over
all four observation epochs.  Clearly, the spectra and luminosity of
the ULXs vary with time.  We discuss this spectral variability further
below.

The best-fitting parameters to the models are in the range $\Gamma
\sim 1.7 - 2.7$ for the PL model, and $kT_{\rm in} \sim 0.8 - 1.7$ keV
for the MCDBB model.  As these two models are also commonly used to
describe Galactic black holes, it is worth considering the comparative
fits gained, although caution should be taken when doing this as the
different band pass of {\it Chandra\/} to those instruments generally
used to study Galactic black holes may be a contributory factor in
differing results (see section \ref{subsection:spectra} for further
discussion). Here we find that the observed values for PL photon index
are in most cases a little on the low side for the steep
power-law state (a.k.a. very high state) as described by McClintock \&
Remillard (2006), which is defined by a $\Gamma > 2.4$ power-law
continuum, and in many cases are similar to the $\Gamma \sim 1.7 -
2.1$ slopes seen in the hard state.  However, this range of power-law
slopes is very typical for {\it Chandra\/} observations of ULXs
(Swartz et al. 2004, Berghea et al. 2008).  In a similar vein, the
MCDBB inner disc temperatures are generally a little high compared to
Galactic black holes in the thermal-dominated (high/soft) state which
generally possess disc temperatures $kT_{\rm in} \sim 0.7 - 1.1$ keV
(although higher temperatures have been seen, particularly in the steep
power-law state cf. McClintock \& Remillard 2006).  They are however
very similar to those measured in other ULXs modelled by a MCDBB
(e.g. Makishima et al. 2000).  The range of spectral properties
therefore appear typical of the ULX class in general.\footnote{An
explicit comparison of our fits to the data from C1, and the published
fits from RWWM02, reveal that our parameter values ($\Gamma, kT_{\rm
in}$) match very closely.  However, there is a trend for our new
absorption values to be $\sim 30\%$ higher (although in many cases
there is reasonable agreement with the previous measurements within
the errors).  This difference is likely attributable to a combination
of considering data below 0.5 keV - which RWWM02 did not - and the
improvements in calibration made over the intervening years.}


\subsubsection{Spectral variability}
\label{subsection:spectral_var}
As the previous section shows, we have multiple detections of the ULXs
in the NGC 4485/4490 system, each with sufficient data quality to
examine their individual X-ray spectra.  This therefore provides an
excellent dataset with which to begin to examine the variation in the
X-ray spectra of these ULXs over time, as amply demonstrated in
Figure~\ref{fig:spectra}.  In particular, five ULXs have a complete
set of spectral fits, i.e. an acceptable - or only marginally rejected
- best-fitting spectrum in each epoch, which we base this analysis on.
We do not discuss either the new transient ULX CXOU J123038.3+413830
(with only one spectral dataset), or the candidate SNR CXOU
J123029.5+413927, in the following work.


\begin{figure*}
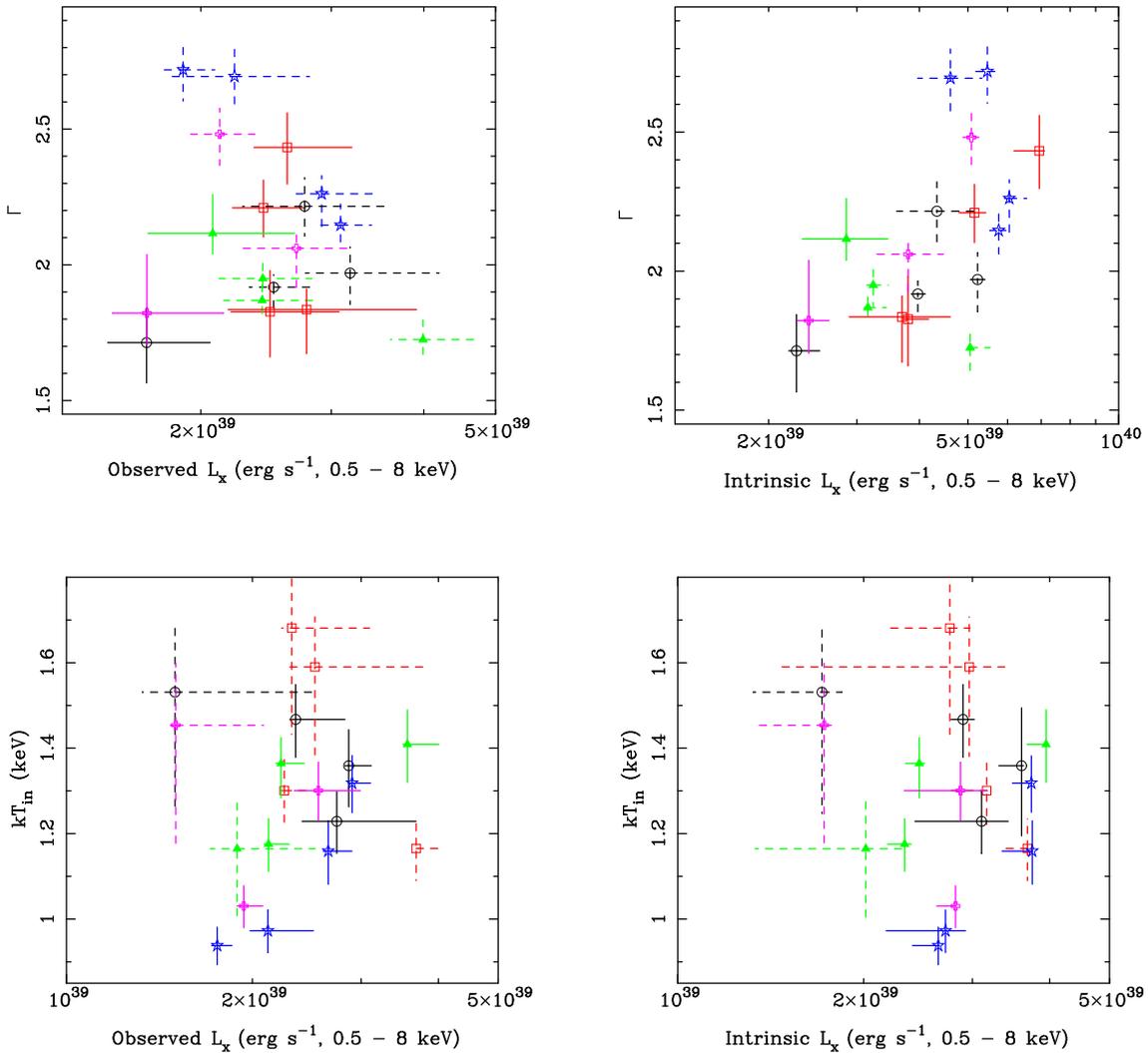

\leavevmode
\begin{center}
\includegraphics[height=70mm, angle=-90]{Lx_v_gamma.ps} \hspace*{1.0cm}
\includegraphics[height=70mm, angle=-90]{Lx_unabs_v_gamma.ps} 
\vspace*{1.0cm} 

\includegraphics[height=70mm, angle=-90]{Lx_v_kT.ps} \hspace*{1.0cm}
\includegraphics[height=70mm, angle=-90]{Lx_unabs_v_kT.ps} 
\end{center}
\caption{Variation of the spectral parameters in five ULXs in the NGC
4485/4490 system.  ({\it Top row\/}) We show the best fitting
parameter for the power-law continuum model, $\Gamma$, against the
0.5--8 keV luminosity derived from that model for two cases: ({\it
left panel\/}) the observed luminosity and ({\it right panel\/}) the
intrinsic luminosity.  ({\it Bottom row\/}) The same, but for the inner
accretion disc temperature $kT_{\rm in}$ in the MCDBB model.  We plot
each pair of panels on the same scale for ease of comparison.  For
ULXs best fitted by the spectral model that is the subject of each
panel, we plot the error bars as solid lines.  Where the ULX is better
fit by the other model, we plot dashed error bars.  Each ULX is
delineated by a different colour/symbol combination.  These are:
green filled triangles - CXOU J123030.6+414142; red open squares -
CXOU J123030.8+413911; black open circles - CXOU J123032.3+413918;
magenta open plusses - CXOU J123036.3+413837; blue open stars - CXOU
J123043.2+413818.
}
\label{fig:specvar}
\end{figure*}

A first interesting trend to consider is which models provide a best
fit to the data. Initially we consider the best fitting model to be
that with the lowest $\chi$$^2$/DoF (although in some instances
$\Delta$$\chi$$^2$ between the models is marginal). On the whole,
these ULXs are more frequently better fit by the MCDBB model than the
PL model (12/20 datasets).  One source is consistently best fitted by
each model; CXOU J123030.8+413911 is always better fitted by the PL
model, whereas CXOU J123043.2+413818 is better fitted by the MCDBB in
the four observations.  The other three can all be best fitted by both
models in one or more observations. However, one trend appears in
these three sources; {\it the PL provides the better fits exclusively
at lower observed luminosity\/}, {\it whereas the MCDBB fits to the
higher luminosity data\/}. An obvious interpretation of this change in
spectral shape is that it could be an indication for a spectral state
transition.  Given the potential importance of this result, we have
investigated it further.  As so few of the spectral fits were rejected
on the basis of the $\chi^2$ statistic, the vast majority of fits
constitute acceptable interpretations of the data so we could not
simply use the $\chi^2$ statistic to distinguish models.  Instead, we
turn to Bayesian Information Criterion analysis (Jefferys 1961;
Schwarz 1978).  This provides a measure of whether one model should be
considered a superior description of the data, when compared to
another.  We calculate values in the range $0.03-1.13$ for the
majority of our comparisons between models for the same epoch data.
This is substantially lower than a value of 2, considered a positive
result by the Bayesian Information Criterion (i.e. evidence that one
model is superior to the other) (cf. Kass \& Raftery 1995).  Hence we
cannot distinguish which model provides a statistically superior fit
to each dataset, so claims that one model clearly provides a better
fit than the other cannot be made on the basis of our data.

Despite the lack of a strong distinction between the two models in
most sources, we do find that the parameterisations of these models
vary between observations.  This is perhaps not unexpected, given that
it is a trend observed in other ULXs (e.g. NGC 2403 Source 3, Isobe et
al. 2008). We have investigated this in Figure~\ref{fig:specvar},
where we plot the derived parameters for each model (the power-law
photon index $\Gamma$, and the inner accretion disc temperature
$kT_{\rm in}$, as listed in Table~\ref{tab:param}) against the 0.5--8
keV luminosity of the source derived from that model.  We show this
for both the observed luminosity, and for an intrinsic luminosity
calculated in the same band by simply setting the absorption acting on
the model to zero.  In each plot we show the data points for which
that model is the better fitting by solid error lines, and use dashed
errors for the worse fitting model.  Each separate ULX is
distinguished via individual colours and symbols (see the Figure
caption for a key).  The Figure is plotted showing $1\sigma$ errors on
the source parameters; clearly there are statistically significant
changes in luminosity and/or parameterisation for each source over the
course of our observations.

What is immediately obvious from comparing the two pairs of panels in
Figure~\ref{fig:specvar} is that the models react rather differently to
the removal of an absorption column.  The removal of an absorption
column for the MCDBB model leads to little movement in the relative
positions of the data points, only a general movement to higher
luminosities (with the latter as would be expected from correcting to
an intrinsic luminosity).  However, the power-law continuum data
points react rather more dramatically, with the data with softer
intrinsic slopes (higher values of $\Gamma$) having a much larger
apparent correction from observed to intrinsic luminosity than those
data with harder slopes.  This is rather well-demonstrated by the data
for CXOU J123030.8+413911 (red squares, the only source always
best-fitted by the power-law model), which appears to vary its
spectral slope with little or no contemporaneous change in observed
luminosity; however when the correction is made for the modelled
absorption, it show a clear trend of becoming softer as its luminosity
increases.  A pertinent question is therefore whether this latter
relationship is a physically correct interpretation of the data, or
whether this is some sort of artefact of the fitting process.

\begin{figure}
\leavevmode
\begin{center}
\includegraphics[height=68mm, angle=-90]{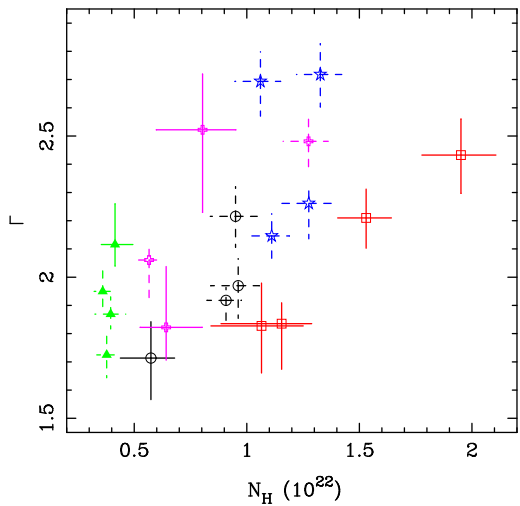} \hspace*{1.0cm}
\end{center}
\caption{ A comparison between the fitted values of the power-law photon index $\Gamma$ and the absorption column $N_{\rm H}$ for the power-law model.  ULXs are denoted by the same colour/symbol combination as in Fig~\ref{fig:specvar}.}
\label{fig:Nh_vs_gamma}
\end{figure}

We have examined this question by plotting the best fitting values of
the power-law photon index $\Gamma$ against the fitted absorption
column density $N_{\rm H}$ in Figure~\ref{fig:Nh_vs_gamma} .
If there were no relation between the two one would see a simple
scatter plot, with the column independent of the power-law slope.
However, at least in the case of CXOU J123030.8+413911, the measured
column density is clearly increasing concurrently with $\Gamma$.  This
suggests that one of two scenarios is occuring: either there is a
physically real increase in absorbing column, local to the ULX, that
occurs simultaneously with a steepening of the spectral slope and increase 
in intrinsic luminosity; or that the spectral fitting process is artificially inferring 
higher columns that go hand-in-hand with steeper slopes in some observations.  
The latter is not so far-fetched; as the column is inferred from the
low-energy turn-over of the power-law spectrum, one might expect some
degeneracy between the slope being turned over and the amount of
material required to produce the turn-over, particularly within the
limited band-pass of the {\it Chandra\/} data where we see few data
points above 5 keV to really constrain the slope of the power-law tail
(cf. Figure~\ref{fig:spectra}).  

The other four sources all behave somewhat differently to CXOU
J123030.8+413911.  Two sources - CXOU J123030.6+414142 and CXOU
J123043.2+413818, plotted in Figures~\ref{fig:specvar} \&
\ref{fig:Nh_vs_gamma} as green triangles and blue stars respectively
- appear to consistently harden their spectra as their luminosity
increases.  Interestingly these two sources are predominantly better
fit by a MCDBB model (albeit with the caveats discussed above about
distinguishing the models), and we note that this behaviour is quite
consistent with accretion discs becoming hotter as their luminosity
increases, as is seen in Galactic black hole X-ray
binaries\footnote{Indeed, standard accretion disc spectra are known to
show a relationship of $L_{\rm X} \propto T^4$, e.g. Done,
Gierli{\'n}ski \& Kubota (2007).}.  On the other hand, the behaviour
of the remaining two sources is less clear cut.  CXOU J123032.3+413918
(plotted as black circles) is mainly better fit by a MCDBB model; in
this model it appears not to vary substantially (less than $\sim
30\%$, with large errors) in luminosity or disc temperature, but when
better fit by a power-law model it appears slightly harder in spectrum
(although again with large errors).  CXOU J123036.3+413837 behaves
similarly - again it appears harder in the power-law state than when
it possesses a MCDBB-dominated spectrum, although in this case there
is a wider difference in the parameters of the data points for the
MCDBB spectrum.  Interestingly, though, these two disc-dominated
spectra appear to follow the same more luminous/harder trend as CXOU
J123030.6+414142 and CXOU J123043.2+413818.  So, although we are
unable to strictly distinguish which model provides the preferable fit
to the data, it does appear as though those sources slightly better
fitted by a MCDBB model are behaving in a fashion consistent with that
expected of sources dominated by an accretion disc.  We discuss this
interesting result further in the next section.

\subsection{Short-term X-ray variability} %

Following the method of RWWM02, light curves for each
source observation were binned to $\sim$25 and $\sim$100 counts per
bin to allow simple tests on variability to be carried out. Initially
these tests were performed by fitting the background-subtracted
light curves to a constant flux (the average for that observation)
then examining the fit with a $\chi$$^2$ test. It was necessary to cut
the (co-added PN and MOS) light curves in two to remove a large
flaring event from the observation, therefore the {\it XMM-Newton}
data initially provides two light curves for each source. As a further
test the {\it XMM-Newton} light curves were reassembled (excluding the
flaring event) and the light curves were re-tested. A separate test on
the light curves was carried out by investigating the excess variance of each dataset.  However, no
statistical signs of short-term variability were detected by any test  in any light
curve.  We note that none of the seven ULXs we have in common with the
work of Fridriksson et al. (2008) showed short-term variability in
their analysis either, when tested using a one-sided
Kolmogorov-Smirnov test.

\begin{figure*}
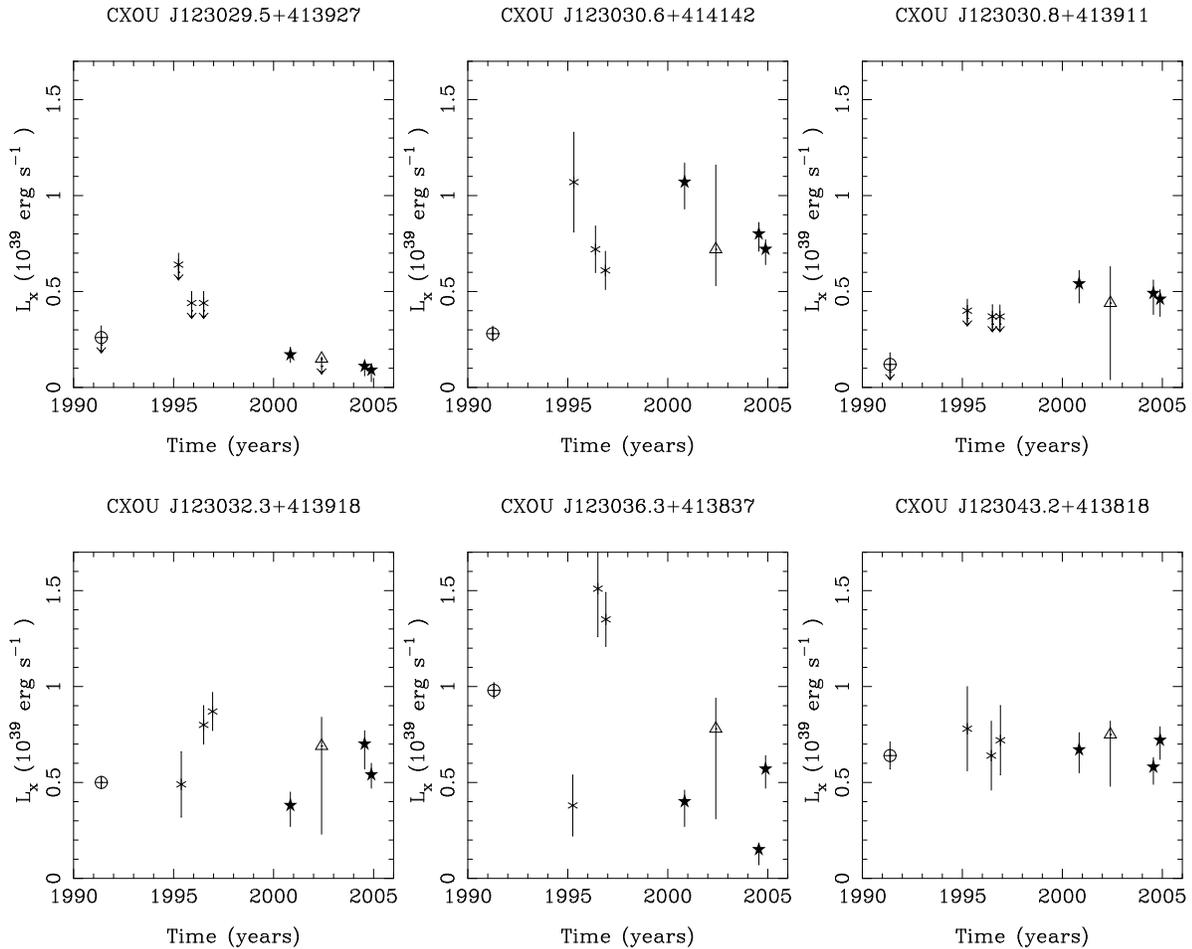

\begin{center}
\leavevmode
\epsfxsize=6cm \rotatebox{-90}{\epsfbox{src4_long_lc.ps}} \epsfxsize=6cm \rotatebox{-90}{\epsfbox{src6_long_lc.ps}} \epsfxsize=6cm \rotatebox{-90}{\epsfbox{src2_long_lc.ps}}\vspace*{5mm} \epsfxsize=6cm \rotatebox{-90}{\epsfbox{src1_long_lc.ps}} \epsfxsize=6cm \rotatebox{-90}{\epsfbox{src7_long_lc.ps}} \epsfxsize=6cm \rotatebox{-90}{\epsfbox{src5_long_lc.ps}}
\end{center}
\caption{Extended long-term light curves of the six ULXs identified in NGC 4485 and NGC 4490. Results shown are from {\it ROSAT} PSPC and HRI instruments (represented by open circles and asterisks respectively), {\it Chandra} ACIS-S (filled pentagram) and {\it XMM-Newton} EPIC camera (open triangle) in the 0.5--2 keV band, using the best fitting single component models (either absorbed power-law or multi-colour disc blackbody). Data from the earlier missions have been taken from the RWWM02 paper, whilst more recent results have been calculated in this work. Upper limits are represented by downwards arrows.}
\label{fig:long term lc}
\end{figure*}

\subsection{Long-term X-ray variability} %

\begin{figure}
\begin{minipage}{85mm}
\begin{center}
\leavevmode
\epsfxsize=6cm \rotatebox{-90}{\epsfbox{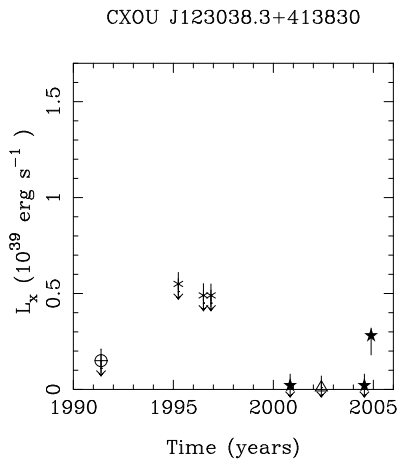}} 
\end{center}
\caption{Extended long-term light curve for the new transient ULX,
CXOU J123038.3+413830.  Symbols are as per Fig.~\ref{fig:long term
lc}.}
\label{fig:src_b lc}
\end{minipage}
\end{figure}

Fridriksson et al. (2008) examine the long-term variability of the
ULXs in NGC 4485/4490, and detect significant flux variability in all
but one of the ULXs (that one being CXOU J123030.8+413911).  Here we
also derive long-term light curves, albeit using generally more
conservative techniques, which we present in Figures~\ref{fig:long term
lc} and \ref{fig:src_b lc}.  We expand on the long-term lightcurves
presented in RWWM02 (based on {\it ROSAT\/} data, and {\it Chandra\/}
data from observation C1 alone), which we limit to the 0.5 - 2 keV
band as this is the only band in common for all the instruments.  The
luminosities we quote are therefore generally lower than seen
elsewhere.  We also include limits where sources are undetected
(calculated as per RWWM02).  Finally, we display the 1$\sigma$ errors for
the luminosities derived from spectral modelling of the {\it
Chandra\/} and {\it XMM-Newton\/} data (the errors on the {\it
ROSAT\/} data are more standard counting errors, due to the lack of
spectral information in this data).

Long-term light curves for the six previously-known ULXs are shown in
Figure~\ref{fig:long term lc}, which is a simple extension of Figure 5 in
RWWM02.  The light curve for the new transient ULX (CXOU
J123038.3+413830) is in Figure~ \ref{fig:src_b lc}, and shows it would
have been clearly detectable in earlier {\it ROSAT\/} PSPC, {\it
Chandra\/} and {\it XMM-Newton\/} data.  Of the other ULXs, CXOU
J123030.8+413911 was itself deemed a transient in RWWM02 due to its
non-detection by {\it ROSAT\/}; however it has apparently remained in
outburst at roughly the same flux level in observations spanning $\sim
4$ yrs after its discovery. The flux of CXOU J123043.2+413818 has
remained similarly stable, though this is over a longer $\sim 15$ yr
timescale. At first glance this may appear to contrast with the findings of Fridriksson et al. (2008), but this difference is predominantly due to bandpass selection. They note significant variations in hard colour that by their definition covers the 1--7 keV band; we however consider the 0.5--2.0 keV band common to all instruments contributing data. To further test our result, we apply the same significance parameter ({\it S}$_{\rm flux}$) used by the authors. They define a source as variable if {\it S}$_{\rm flux}$ $>$ 3. We find that {\it S}$_{\rm flux}$ $=$ 1.3 for CXOU J123043.2+413818, confirming the lack of variability. The three other brighter ULXs have displayed higher variability amplitudes over the $\sim 15$ yrs covered by the observations, though have remained detectable throughout this time (CXOU J123030.6+414152, {\it S}$_{\rm flux}$ $=$ 6.2; CXOU J123032.3+413918, {\it S}$_{\rm flux}$ $=$ 3.6; CXOU J123036.3+413837, {\it S}$_{\rm flux}$ $=$ 12.2). We discuss the remaining source (CXOU J123029.5+413927) below in section 4.2.

\section{Discussion}

\subsection{Spectral state changes in ULXs?}
\label{subsection:spectra}

In the previous section we have considered the properties of the ULXs separately.  We show that most are adequately fitted by two simple spectral models (a power-law continuum or a MCDBB).  Although there are apparent trends (based on a superior $\chi^2$ fit) in the best fitting model to the data with the luminosity of the ULX at a given epoch, we show that statistically we cannot distinguish a best fitting model on this basis alone.  Interestingly, though, the best fitting parameters do give some support to the initial distinction based on the $\chi^2$ fit - those models better fitted by a MCDBB do appear to behave in a way consistent with this (increasing in disc temperature as their luminosity increases).  No temporal variability is evident on short (intra-observation) timescales for any source contained within NGC 4485/4490.  However, strong variability is seen over longer ($\sim$ years) time scales in most of the ULXs.

When we consider the spectral and temporal behaviour of the ULXs together, we find that these ULXs appear rather heterogeneous, an unsurprising result given previous multi-epoch observations (e.g. Fabbiano et al. 2003a, Roberts et al. 2004, Zezas 2007).  We initially consider only the five sources discussed in section \ref{subsection:spectral_var}, examining both the long-term light curves and spectral variability of each source to see if and how they are related. We find that the two sources remaining constant in their long-term light curves display very different spectral evolution behaviours over time. We find that CXOU J123043.2+413818 follows a similar spectral evolution to a disc, loosely following $L_{\rm X}$ $\propto$ $T^4$ on the parameter-luminosity plot, whilst CXOU J123030.8+413911 appears to show the opposite trend,  with $kT_{\rm in}$ decreasing as the luminosity increases. Although we cannot statistically separate model preference, it is interesting to note that each source is consistently better fitted by a particular model.  While CXOU J123043.2+413818  varies in a disc-like manner, it also retains a lower $\chi$$^2$ fit for MCD, whilst CXOU J123030.8+413911 is better fitted by a PL-type shape. We reiterate that we cannot offer a statistical distinction between these two models for either source, but suggest that this secondary evidence supports the initial distinction made by $\chi^2$ fits.

More interestingly, we find that the spectral fits to the other three sources, CXOU J123030.6+414142, CXOU J123032.3+413918 and CXOU J123036.3+413837, appear to show them transiting between these two models, which could potentially be used as a test of our secondary evidence. If we consider their changing spectral behaviour we find that those observations better modelled by a MCDBB appear to vary in a disc-like fashion in the majority of cases, while those modelled by a PL tend to leave the $L_{\rm X}$  $\propto$ $T^4$  track. We also note that the long-term light curves of each of these sources vary substantially; crucially these flux variations correlate with apparent spectral state, with observations at low luminosities appearing to possess a PL-type shape, with spectra switching to a MCDBB-type state at higher luminosities.  Again, despite the lack of statistical evidence for a distinction between the models, this behaviour does appear to support the $\chi^2$ fitting.  If we do accept the changes in spectral shape, then the most obvious explanation of the behaviour of the ULXs is that they can reside in different spectral states, delineated by the PL and MCD spectra, and that in some cases we are observing sources that have undergone a transition from a PL-Type to MCD-type state (or vice-versa). As this is an extremely interesting result, we now consider the physical implications of such behaviour on our understanding of these systems. 

A further reason the think that there may be some physical basis for the two spectral types we see in the NGC 4485/4490 ULXs is that they have been seen in other ULXs before.  Mizuno et al. (2007) suggested (based on observations of two ULXs in NGC 1313) that the PL-type state may be representative of the very high state observed in Galactic black hole X-ray binary systems, whilst the MCDBB-type state may be more suggestive of the theoretically predicted ``slim disc'' model (e.g  Watarai et al. 2001). The latter is supported by observations fit by the $p$-free model (e.g. Vierdayanti et al. 2006). The $p$-free model assumes that the disc temperature scales as $r^{-p}$, where $r$ is the radius and $p$ is a free parameter. In the case of a standard disc the value of $p$ would be fixed at 0.75 but if the $p$-value decreases, softer energies become more enhanced, and the spectrum becomes more representative of a slim disc. In the case of the MCDBB-type spectra of M81 X-9, the best fitting parameters reduce the $p$-value to 0.6, significantly smaller than a standard disc value (Miyawaki et al. 2006), supporting the existence of an ultraluminous slim disc state. Kuncic et al. (2007) made similar associations for the PL-type state, but suggested that this could also be representative of a system with some kind of outflow. The authors did this by demonstrating that a disc modified with jet emission could also explain the observed spectral shape. Soria \& Kuncic (2008) developed this work further, presenting the idea that the PL-type state could be explained by a black hole in the hard state with an associated jet that persisted up to $\sim$ Eddington rates, where an outflow-dominated or a slim disc state would be formed. This allows for the possibility of direct transition from a hard PL-type state to a super-Eddington state.

The suspected change in state that we observe is consistent with those found previously and we argue that, although the PL-like spectrum observed appears similar to that of an accreting black hole in the low/hard state, we are probably observing sources residing in the very high state (VHS; a.k.a. steep power-law state, McClintock \& Remillard 2006). The spectrum of the VHS is empirically characterised by a hot disc and a steep power-law in the 2.0--20.0 keV band. It should be noted that the power-law component is used as a proxy to describe optically-thin Comptonisation, however a spectrum resulting from this process is subtly curved with a shallower slope at lower energies. Here we observe only the lower 0.3--10.0 keV energy range, although the relatively poor high energy sensitivity of the {\it Chandra} ACIS-S means we only have constraints up to $\sim$5 keV. Hence we observe only the shallower section of the Comptonised component.  This should then also be combined with emission from the 1--2 keV disc component seen in the VHS (although data quality is insufficient at present to separate these components). This amalgamation of components could potentially act to flatten the slope of the spectrum below 5 keV, resulting in the range of power-law photon indices observed. As the mass accretion rate (and so the observed luminosity) is increased, it is likely that a change in the accretion structure of the system occurs that invokes a change in spectrum to the MCDBB-like state. The observed curvilinear structure of such a state would be representative (in our modest spectral quality regime) of a slim disc, or an optically-thick Comptonising corona masking the inner radii of the disc (Done \& Kubota 2006; Stobbart et al. 2006). We note that in each of these model scenarios the disc temperature increases with luminosity (Roberts et al. 2006; Vierdayanti et al. 2008).

The detection of this possible change in state (from PL to MCDBB-type or vice-versa) has previously been noted in other ULX studies. The clearest transition to date was reported in M81 X-9 by Miyawaki et al. (2006), with a transitional luminosity of 1.5$\times$10$ ^{40}$ erg s$^{-1}$.  Multiple spectra of NGC 1313 Source B were also analysed by the authors but results in this case were not as clear as M81 X-9. The tendency towards MCDBB-like spectra with increasing luminosity was apparent, although a precise transition was not evident, occuring within the range  $\sim 5 - 9 \times$10$^{39}$ erg s$^{-1}$. Similar transitions, from a PL-type state to a MCDBB-type state, have also occured in IC 342 X-1 and X-2 at around 10$^{40}$ erg s$^{-1}$ (although with only two observations analysed, it is impossible to give a precise transition luminosity; Kubota et al. 2001).

Soria \& Kuncic (2008) proposed that the transition between states would occur at around the Eddington limit, if we assume little beaming. Based on this proposition the observations reported here provide a potential diagnostic to these systems. allowing us to place constraints on the masses of the black holes contained within the ULXs in NGC 4485/4490. Firstly we note that it is certainly not necessary to invoke an intermediate mass black hole to achieve the observed luminosities of these systems ($\la$ 4$\times$10$^{39}$ erg s$^{-1}$), indicating that these sources need be no more than large stellar-mass compact objects ($\la 30 M_{\odot}$ if they obey the Eddington limit).  Secondly, if this apparent state change is occurring at the Eddington limit, the luminosity at which a transition is observed would give a rough estimate on the mass of the system. Although we cannot observe a distinct change in state at this level of data quality, the variable ULXs in NGC 4485/4490 appear to change state at $\sim$ 2$\times$10$^{39}$ erg s$^{-1}$, which would mean that these systems are host to $\sim$10--15 M$_\odot$ black holes, i.e. standard stellar-mass black holes.  Following this argument CXOU J123030.8+413911, which continues to reside in the PL-type state up to a luminosity of 2.8$\times$10$^{39}$ erg s$^{-1}$, could therefore contain a more massive black hole. Interestingly, if we were to extrapolate this argument to other sources where transitions have been reported, with observed transitional luminosities ranging from  2--15$\times$10$^{39}$ erg s$^{-1}$ (including our current sample), the primaries of ULXs would range in mass up to no more than $\sim$100 $M_\odot$ (M81 X-9, in agreement with Tsunoda et al 2006). 

The implications of this argument are in agreement with evidence emerging from other observations of ULXs, a review of which was compiled by Roberts (2007). For example, when high quality X-ray spectra are available, a break emerges above 2 keV (Stobbart et al. 2006), which should not be present in the X-ray spectrum of large, $\sim 1000 M_{\odot}$ intermediate-mass black holes. This feature is more indicative of some form of extreme accretion onto a smaller compact object. Optical evidence also appears to be converging on similar mass estimates, for example irradiation models describing the optical emission of ULXs have been used to constrain the mass of the black hole for several sources, generally imposing the limit M$_{BH}$ $\la$ 100 M$_\odot$ (Copperwheat et al. 2007). Recent developments in theory have shown that it is possible to create black holes with  M$_{BH}$ $\la$ 100 M$_\odot$, in young stellar populations with either low metallicity stars (Fryer \& Kalogera 2001, Heger et al. 2003), or with the merging of binaries (Belczynski et al. 2006). With evidence from both observation and theory converging on large stellar-mass objects, timely evidence came when Prestwich et al. (2007) found such an object. Repeated optical exposures of the Wolf-Rayet black hole binary known as IC10 X-1 culminated in a radial velocity curve, providing a primary mass estimate of 23--34 $M_\odot$, the largest stellar-mass black hole found to date (see also Silverman \& Filippenko 2008).  Hence it is looking an increasingly realistic proposition that a large fraction of the ULX population are powered by black holes not substantially larger than those we already know of in our own Galaxy.

\subsection{The candidate SNR CXOU J123029.5+413927}

CXOU J123029.5+413927 was identified as a possible supernova remnant due to its high absorption column, soft underlying continuum and its coincidence with the radio source FIRST J12309.4+413927 (RWWM02). Mid-infrared studies of NGC 4485/4490, carried out by V{\'a}zquez et al. (2007) using the {\it Spitzer Space Telescope}, found that five of the ULXs showed high ionization features that are found in AGN. Conversely the mid IR spectrum of CXOU J123029.5+413927 showed emission more indicative of star forming regions. This, with the addition of a weakly detected [O\textsc{VI}] 25.91 $\mu$m emission line, a line that is recognised as a strong feature in SNRs (Morris et al. 2006; Williams et al. 2006), supports the previous classification.  If it is a SNR then it is obviously amongst the brightest of its class - X-ray luminosities in excess of $10^{39}$ erg s$^{-1}$ are seen in a number of young SNRs (Immler \& Lewin 2003), although typical older SNRs are 2 - 3 orders of magnitude fainter in X-rays (e.g. Pannuti et al. 2007).  However, in contrast to previous evidence, analysis by Fridriksson et al. (2008) reported a 30\% drop in luminosity for this source in the four months between the second and third {\it Chandra} observations, behaviour that is more typical of an X-ray binary source. 

A compelling argument on the nature of this source is discovered when looking at the light curve of this source. We find that  previous observations place its detonation (or at least the epoch at which its X-ray emission brightened) some time after the 1991 {\it ROSAT} observation, which would imply that we may be observing a very young SNR.  The calculated rate of decay of X-ray flux within the light curve gives further insight. Studies performed using data from early X-ray telescope missions showed that this decay proceeds as $L_{\rm X} \propto t^{-s}$ with index $s = 1$ (e.g. Chevalier \& Fransson 1994); however the flux and temperature calculations performed in these earlier studies could have been contaminated by diffuse emission due to the poorer resolution of these satellites. More recent studies with {\it Chandra} and {\it XMM-Newton} have demonstrated that the decay observed within the light curve of a young SNR has an index $s \sim 3$ (e.g. $L_{\rm X}\propto$ t$^{-2.7}$ in Immler \& Kuntz 2005, t$^{-3}$ in Temple, Raychaudhury \& Stevens 2005, t$^{-3.9}$ in Aretxaga et al. 1999). The rate of decay observed in the light curve of CXOU J123029.5+413927 is found to be  $L_{\rm X}\propto t^{-0.3}$, a decay rate that is far shallower than young SNRs.  Therefore this source {\it could} simply be an X-ray binary system, as suggested by Fridriksson et al. (2008). An alternative to this could be a young SNR with an additional roughly constant component, diluting the emission and affecting the observed decay rate. This component could be an X-ray binary system, although it is notable that it hasn't produced the high ionisation nebulae as seen around the other ULXs in NGC 4490 by V{\'a}zquez et al. (2007).  Much deeper observations would be required in order to test either hypothesis.  

\section{Conclusions}

Multiple observations of the interacting galaxies NGC 4485 \& 4490 have afforded us the opportunity to study its large ULX population, and in particular to see how their emission evolves over time. We have found that the population is unanimous in its lack of short-term temporal variability, but long-term spectral and temporal variations have given a possible insight into the mass of black holes contained within them. Observations have revealed that sources exhibiting large scale temporal variation (excluding CXOU J123029.5+413927) may change spectral shape from that of a PL-type to a MCDBB-type as their luminosity increases. If real - and there is some uncertainty here, as we cannot strongly differentiate between fits with PL and MCDBB models based on the current data -- this change in spectral shape/state occurs at $\sim 2 \times 10^{39}$ erg s$^{-1}$ for the ULXs in this system. We propose that this state change is from the VHS to an ultraluminous state, a transition that occurs at around the Eddington limit. If this is the case it would imply that {\it the majority of the ULX population residing in this interacting galaxy pair are stellar-mass black holes of around 10--15 $M_\odot$}.

The analysis of the possible spectral transitions has culminated in a testable hypothesis that can potentially be applied to other ultraluminous X-ray sources, for example the newly observed transient source CXOU J123038.3+413830. This object has only been observed once to date and so it is impossible to tell at present any more than its current preferred state, a MCD-type state. The observed luminosity of this source is much lower than the transitional luminosity noted for the majority of sources in our current source population. Extrapolating from our above arguments would suggest that this system contains a compact object of lower mass. Further observations are obviously required to constrain the nature of this source and indeed to continue our study of the behaviour of the whole ULX population of NGC 4485/4490.  Fortunately, this will be possible using time allocated to us on {\it XMM-Newton\/}.  The resulting data set -- to be reported in a follow-up paper -- will help us delve deeper into the nature of the remarkable ULX population residing in these galaxies.

\section*{Acknowledgements}

We thank the anonymous referee for their constructive comments, that have helped to improve this paper.  JCG gratefully acknowledges funding from the Science and Technology Facilities Council (STFC) in the form of a PhD studentship. This work is mainly based on data from the {\it Chandra} satellite, which is operated by the National Aeronautics and Space Administration (NASA).  It is also partially based on observations obtained with {\it XMM-Newton\/}, an ESA Science Mission with instruments and contributions directly funded by ESA member states and the USA (NASA). The Digitized Sky Surveys were produced at the Space Telescope Science Institute under U.S. Government grant NAG W-2166. The images of these surveys are based on photographic data obtained using the Oschin Schmidt Telescope on Palomar Mountain and the UK Schmidt Telescope. The plates were processed into the present compressed digital form with the permission of these institutions. The National Geographic Society - Palomar Observatory Sky Atlas (POSS-I) was made by the California Institute of Technology with grants from the National Geographic Society. The Second Palomar Observatory Sky Survey (POSS-II) was made by the California Institute of Technology with funds from the National Science Foundation, the National Geographic Society, the Sloan Foundation, the Samuel Oschin Foundation, and the Eastman Kodak Corporation. The Oschin Schmidt Telescope is operated by the California Institute of Technology and Palomar Observatory. The UK Schmidt Telescope was operated by the Royal Observatory Edinburgh, with funding from the UK Science and Engineering Research Council (later the UK Particle Physics and Astronomy Research Council), until 1988 June, and thereafter by the Anglo-Australian Observatory. The blue plates of the southern Sky Atlas and its Equatorial Extension (together known as the SERC-J), as well as the Equatorial Red (ER), and the Second Epoch [red] Survey (SES) were all taken with the UK Schmidt. Supplemental funding for sky-survey work at the ST ScI is provided by the European Southern Observatory. 


\label{lastpage}

\end{document}